# On-chip detection of spin-selective routing in plasmonic nanocircuits

*Martin Thomaschewski,\* Yuanqing Yang,\* Christian Wolff, Alexander S. Roberts, and Sergey I. Bozhevolnyi*

Center for Nano Optics, University of Southern Denmark, Campusvej 55, DK-5230 Odense M, Denmark

*Corresponding Authors: math@mci.sdu.dk, yy@mci.sdu.dk

**ABSTRACT** On-chip manipulating and controlling the temporal and spatial evolution of light is of crucial importance for information processing in future planar integrated nanophotonics. The spin and orbital angular momentum of light, which can be treated independently in classical macroscopic geometrical optics, appear to be coupled on subwavelength scales. We use spin-orbit interactions in a plasmonic achiral nano-coupler to unidirectionally excite surface plasmon polariton modes propagating in seamlessly integrated plasmonic slot waveguides. The spin-dependent flow of light in the proposed nanophotonic circuit allows on-chip electrical detection of the spin state of incident photons by integrating two germanium-based plasmonic-waveguide photodetectors. Consequently, our device serves as a compact (~ 6×18 μm$^2$) electrical sensor for photonic spin Hall dynamics.

The demonstrated configuration opens new avenues for developing highly-integrated polarization-controlled optical devices that would exploit the spin-degree of freedom for manipulating and controlling subwavelength optical modes in nanophotonic systems.

**Introduction.** Light carries both the spin, an intrinsic form of angular momentum, and orbital angular momentum, which determines its polarization and spatial degree of freedom. Interaction between the spin and orbital degrees of freedom of photons has evoked intensive investigations owing to its potential to push the development of technologies, such as chiroptical spectroscopy[1,2], communication[3], information processing[4], topological photonics[5,6] and quantum computing[7], to their full potential. The limiting factor for groundbreaking developments in those fields refers to the fact that the spin-orbit interactions (SOIs) in optics are usually very weak, akin to the Planck-constant smallness of the electron SOI found in solid-state spintronics[8]. A promising way to significantly enhance spin-controlled optical phenomena is to utilize light-matter interactions on the nanoscale that are especially strong in plasmonic nanostructures. It has been shown that geometrically chiral metallic structures, which do not superimpose onto their mirror image, can strongly enhance chiroptical far-field responses as a consequence of structural chirality[9–12]. Remarkably, even achiral structures exhibit the SOI potential in the near-field due to twisted trajectories of surface plasmons at a nanosphere[13–16]. This feature enables spin-controlled local manipulation within one nanoscale coupler, which responds equally to both photonic spin states. We utilize the strong SOI in an achiral plasmonic nanostructure to demonstrate for the first time on-chip detection of spin-controlled directional routing in a compact plasmonic nanocircuit. We find that a subwavelength semiring can launch gap surface plasmons supported by seamlessly integrated plasmonic slot waveguides preferentially in one direction, depending on the spin state of locally incident radiation. This spin-dependent phenomenon can thus be regarded as a manifestation of the quantum spin Hall effect (QSHE) of light[5,8]. We bridge the fields of integrated photonics and electronics, progressing towards what has repeatedly been highlighted as the key perspective of plasmonics[17–21], by integrating two germanium-based plasmonic photodetectors for on-chip electrical read-out of unidirectional surface plasmon polariton (SPP) excitation. We show that non-chiral optical responses, which are inherently dependent on the linear polarization degree, can be quantitatively identified and discriminated based on synchronous differential photocurrent detection, providing a signal directly proportional to the $S_3$ Stokes parameter of locally incident radiation. Our fabricated device thus serves as a miniature chiral light detector, a functionality that is generally difficult to realize with conventional photodetectors[22]. In addition, the same structure offers the ability of identifying linear polarization states, making it particularly interesting for fast polarimetric imaging.

**Results and Discussion.** Figure 1 schematically shows the proposed plasmonic device, which provides the features mentioned above. It comprises a non-chiral gold semiring waveguide coupler with a width of 150 nm and a radius of curvature of 225 nm in which normally incident, focused light is preferably exciting SPPs in one of the two seamlessly integrated gold metal-isolator-metal (MIM) slot waveguides due to SOI. The excited propagating mode is strongly confined in the sub-diffractional waveguide slots, which have sizes of w × h = 150 × 100 nm². The entire device is covered with a 300 nm-thick dielectric layer (PMMA, $n_{PMMA}$ = 1.48) to reduce mode leakage into the underlying $SiO_2$ substrate. To prevent adiabatic coupling between the two slot waveguides, we separate the channels along the SPPs propagation direction. Using this design, plasmonic photodetectors are integrated by locally filling the slot with germanium (Ge) semiconductor material in each individual branch. The strongly confined SPP mode is absorbed by the Ge material and the generated electron-hole pairs are collected by an external electric field. The inset in Figure 1a provides a schematic representation of the energy band diagram of the Au-Ge-Au heterostructure while the detector is biased with an externally applied voltage, which attracts the carriers towards the metal contacts. The electrodes of the plasmonic photodetector are naturally integrated by the metal that supports the propagation of the corresponding SPP mode. For interfacing the electrical nanocircuit to macroscopic electrodes, three connecting wires are embedded, allowing for independent operation of two detectors by individual bias electrodes with respect to a common ground. In this way, fabrication imperfections that might result in unequal photo-responses of the detectors and/or unequal propagation losses within the two waveguide branches can be easily compensated by a single adjustment of the reverse bias that defines the carrier extraction efficiency of each individual detector. Consequently, the entire device allows for differential photocurrent measurements of two plasmonic photodetectors, which directly reveal accurate information about the spin state of the locally incident beam. Fabrication details of the proposed device are given in the 'Methods' section.

In order to study the SOI in our device, we first focus on the interaction of incident radiation with the semiring coupler. Therefore, we validate numerically the spin-sorting functionality by utilizing finite difference time-domain (FDTD) simulations, considering that the modeled geometry is illuminated with a diffraction-limited beam ($\lambda_0$ =1550 nm). We clearly see in Fig. 2a that under left-circular polarization (LCP) and right-circular polarization (RCP) illumination, the field is concentrated on one side of the looped wire, respectively, and the excited plasmons stay strongly localized at the corresponding waveguide resulting in unidirectional power flow. The origin of the unidirectional excitation can thus be understood by consideration of the twisted near-field evolution within the semiring. While the majority of the incident light is coupled into one particular waveguide, a small fraction of optical power is directed into the opposite waveguide, leading to a theoretical directionality contrast of 14.5 dB. To gain a deeper insight into the excitation mechanisms governing the Poynting vector flow, the directed power in the individual channels is studied for various polarizations ranging from linear to circular with intermediate elliptical polarization states (See Fig. 2b). We observe that any linear polarization excites both waveguides equally due to equal contributions from LCP and RCP. However, the total power flow is enhanced at a linear polarization perpendicular to the waveguide. This observation indicates that the system exhibits a non-chiral excitation mechanism, which couples light simultaneously into both slot waveguides once the polarization coincides with the antisymmetric slot waveguide modes[23]. As the power flow is governed by two independent coupling mechanisms, the maximum of the directed intensity appears at elliptical polarization, slightly shifted from circularly polarized states towards vertical linear polarizations.

With the above numerical simulations establishing the framework for spin-selective unidirectional excitation, we now move on to the experimental realization of the on-chip device for spin-selective routing. For this purpose, we first conduct an all-optical far-field characterization of the device before plasmonic photodetectors are integrated (See Figure 3). A diffraction-limited beam with wavelength $\lambda_0$ =1550 nm and controlled polarization is positioned onto the coupler section of the device at normal incidence. By adjusting the retardance of a Soleil-Babinet compensator, we generated a beam with LCP and RCP. Light scattering from the impedance-matched nanoantennas is observed in reflection mode for both polarization states (Figure 3b and 3c). The emission intensities at the impedance-matched nanoantennas reveal that the plasmonic modes generated by photons with opposite spin are efficiently directed to different waveguide branches.

After experimentally verifying the predicted spin-sorting functionality of the device, plasmonic photodetectors are integrated for the realization of full on-chip detection. Therefore, an additional lithography step was carried out to define the plasmonic detector region. Consequently, the plasmonic slot waveguide is locally filled with amorphous Ge semiconductor material by thermal evaporation. To substantiate the electrical detection of propagating plasmons, we have comprehensively characterized a plasmonic photodetector in a simplified system from the same fabrication batch, consisting of a straight plasmonic slot waveguide (See Fig. S1, Supporting Information). In our device, the launched plasmonic mode propagates towards the detector and penetrates the MSM region. The SPPs are absorbed in the semiconductor material generating electron-hole pairs which are efficiently separated by an externally applied electric field $E \sim U/w$. Internal quantum efficiencies (IQE) exceeding 1% are demonstrated for the individual detector at the targeted wavelength of 1550 nm.

For the on-chip electrical detection experiment, the beam is positioned onto the coupler section and the photocurrents of the individual photodetectors are measured for various polarization states of the incident beam. Figure 4 shows measured photocurrents with its differential signal, where the polarization angle, ellipticity and handedness of the incident beam is continuously varied. It is seen that the SOI process in our device is directly translating the spin-controlled contribution to the coupling process into unbalanced photocurrents of the two detectors. In contrast, any linear polarization state shows no directionality due to a symmetric optical coupling process, resulting in a near-zero signal in the differential photocurrent measurement (Figure 4b). This reveals the fact that directionality in our device is exclusively induced by SOI. Considering non-linear states of polarization, our device has the crucial feature of providing a differential photocurrent signal which can be directly translated to the handedness and helicity of the incident beam, while being virtually insensitive to linearly polarized light (Figure 4a). Non-chiral optical effects inherently cancel out so that the sign of the corresponding signal displays the handedness and the absolute value its helicity. We find that the directionality appears with non-resonant characteristics which enables broadband operation (See Supporting Information, Section S2). Comparing the absolute photocurrent evolutions of the individual detectors with the numerical investigation in Fig.3 indicates that non-chiral excitation mechanisms are more pronounced in the experiment than predicted by simulation. We believe that this discrepancy is associated with fabrication imperfections and surface roughness in the semiring and waveguides which serves potentially as inherent waveguide coupler due to diffraction on roughness features.

**Conclusion.** In summary, using SOI in an achiral semiring, electrical on-chip detection of spin-selective directional routing in a plasmonic nanocircuit has been demonstrated successfully. This study may open new routes in highly integrated plasmonic nanocircuits that include the spin degree of freedom for manipulating the optical power flow, in analogy to electron spintronics. Our proposed device not only enables spin-controlled tunable addressing of two plasmonic waveguide channels for spin-encoded communication applications, but also serves as a compact polarimeter capable of determining the one-axis linear polarization state, helicity and handedness of locally incident radiation. The device can be extended by waveguide channels that are sensitive to linear polarization along one of the ±45º axes with respect to the symmetry axis of the device, which enables a complete on-chip characterization of the polarization state. Taken together, we believe this study further bridges the remarkable nanophotonic manipulation capability of plasmonics with nanoscale electronic systems.

**Methods.** *Fabrication.* The fabrication of the device relies on a multi-step lithographic process using the mix-and-match technique for lithographic overlay. First, bonding pads and connecting electrodes are patterned onto a glass chip by optical lithography, metal deposition (5 nm Ti / 50nm Au), and lift-off. The plasmonic circuit is written by electron-beam lithography (EBL) at an acceleration voltage of 30keV in 300 nm PMMA resist and 20 nm thick Al layer to prevent charge accumulation. The alignment between the different lithography steps is performed manually using markers. After development, the circuits are formulated by depositing a 5 nm titanium adhesion layer and 100 nm gold by thermal evaporation and subsequent 12 hours lift-off in acetone. For the optical characterization a 300 nm PMMA cladding was spin-coated on the chip to reduce mode leakage into the glass substrate. The same PMMA layer serves as an electron beam resist for the last EBL step, which integrates the germanium photodetectors onto the plasmonic circuit by evaporating 2 nm Ti and 110 nm Ge onto the pre-defined region. Finally, a 300-nm thin layer of PMMA is deposited. For electrical connection, the chip is wire bonded on a printed circuit board equipped with two on-board amplifiers.

***Numerical Simulations.*** Finite-difference time-domain (FDTD) simulations were performed using a commercial software (FDTD Solutions v8.11.318, Lumerical). Perfectly matched layers (PMLs) were applied to enclose the computational domain of 3 μm × 4.5 μm × 2 μm in all calculations. A finest mesh grid size of 10 nm is used in the simulations. The optical constants of the gold were taken from the experimental data reported by Johnson and Christy[24]. For circular polarization excitation, we employed two coherent Gaussian sources with the same intensity but orthogonal polarizations with a phase retardance of 90°. The diameter of the Gaussian beam was set to a 2 μm. To determine the coupled light into each waveguide, we integrated the time-averaged Poynting vector over the slot gap area at 3 μm away from the antenna coupler.

AUTHOR INFORMATION

Corresponding Authors

*E-mail: math@mci.sdu.dk, yy@mci.sdu.dk

Notes

There are no conflicts of interest to declare.

ACKNOWLEDGMENT

The authors would like to thank for financial support from the European Research Council (Grant 341054, PLAQNAP), the VILLUM FONDEN (Grant 16498) and the University of Southern Denmark (SDU2020 funding). M.T. acknowledges advice from Jakob Kjelstrup-Hansen on the micro- and nanofabrication steps.

AUTHOR CONTRIBUTIONS

Y.Y. conceived the idea and performed the simulation. Y.Y. and M.T. designed the final device structure. M.T. set up the experiment and performed the device fabrication and experimental measurements. C.W. and A.S.R assisted in the experimental measurements. All authors contributed to the interpretation of the results. M.T. wrote the manuscript with contributions from all other authors. S.I.B. supervised the project.

**Figure 1**

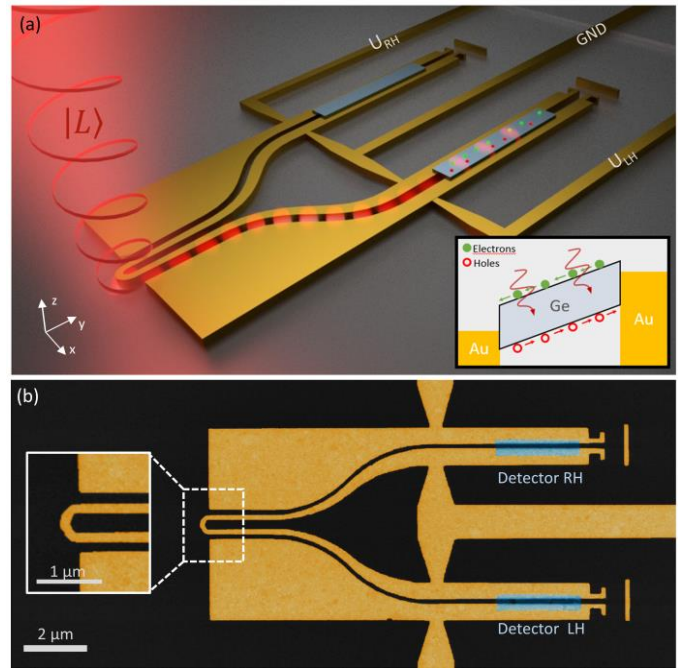

***Figure 1*** *Schematics and design of the proposed device utilized for on-chip electrical detection of spin-controlled unidirectional plasmonic waveguiding. (a) Three-dimensional rendering illustrates exemplarily for a left circular polarized incident beam the spin-selective unidirectional routing in the plasmonic nanocircuit. The propagating waveguide mode is reaching the plasmonic metal-semiconductor-metal (MSM) photodetector consisting of a Au-Ge-Au heterostructure, in which electron-hole pairs are generated and subsequently collected by an externally applied reverse bias. Inset: energy band diagram of the Au/Ge/Au heterojunction MSM photodetector under bias. (b) Colorized scanning electron microscopy image showing the top view of a plasmonic device. The Ge region (azure color) on top of the slot plasmonic waveguide forming the plasmonic detector can be readily observed in the SEM image.*

**Figure 2**

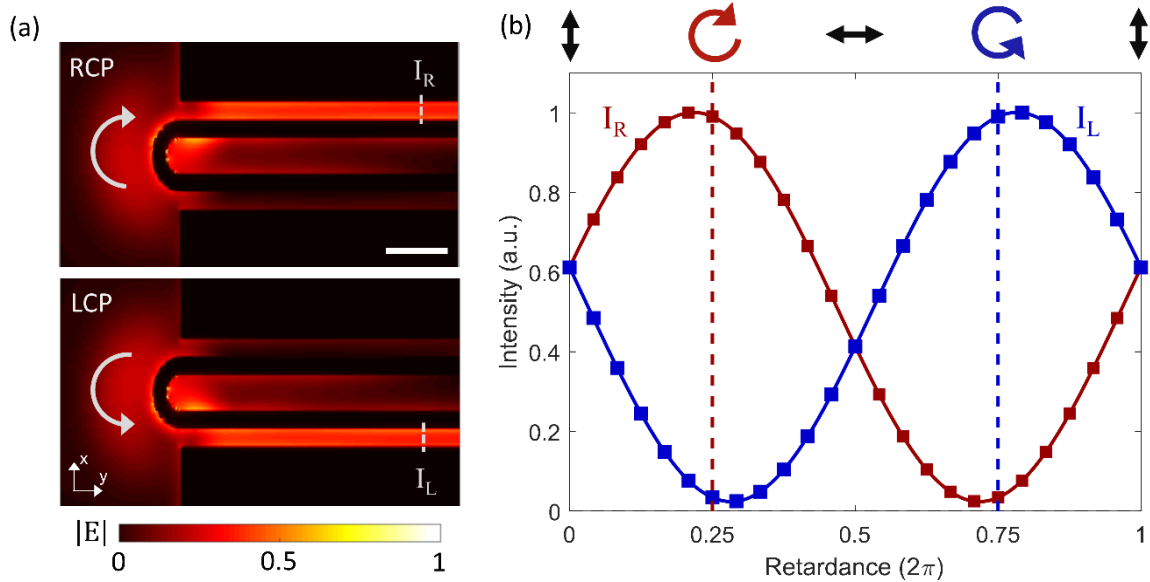

*Figure 2* Nanoscale directional coupler enabling selective far-field excitation of two plasmonic slot waveguides by spin-orbit interaction. (a) Simulated near-field intensity distribution for excitation with LHC and RHC input polarization demonstrating routing to the left and right branch, respectively. The scale bars represent 500 nm. (b) Output intensities of the two channels at continuously varying retardation between the polarization axis perpendicular and parallel to the waveguide axis. 0 and $2\pi$ retardance correspond to linear polarization perpendicular to the waveguides axis, while $\pi$ correspond to linear polarization parallel to the waveguide. In-between, circular polarized light with opposite spin are present.

**Figure 3**

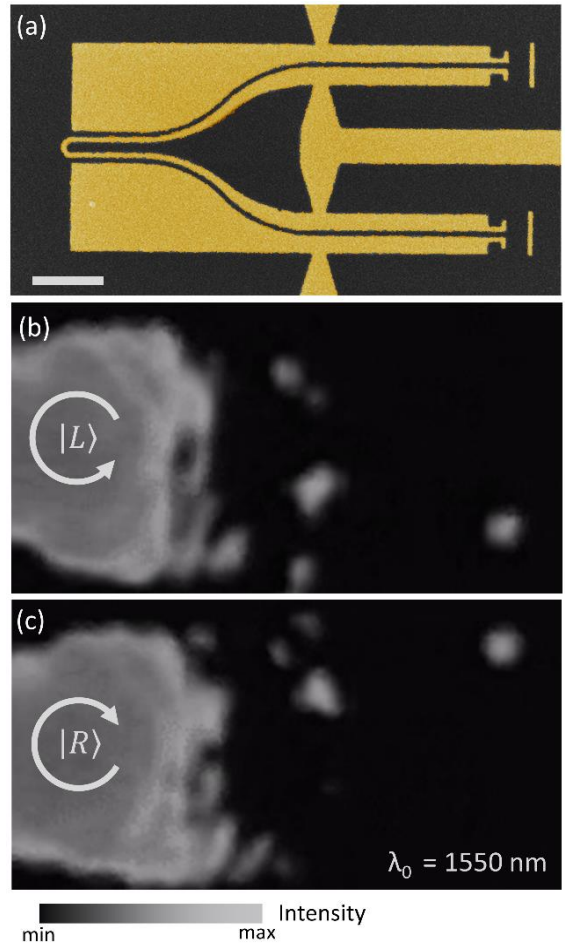

***Figure 2** All-optical experimental demonstration of spin-sorting in the proposed plasmonic nanocircuit. (a) Colorized scanning electron microscopy image of the device before plasmonic photodetectors are integrated. The scale bars represent 2 μm. (b-c) The appearance of the far-field scattering signal at different antennas reveals the fact that photons with opposite spin are efficiently directed into opposite waveguide branches.*

**Figure 4**

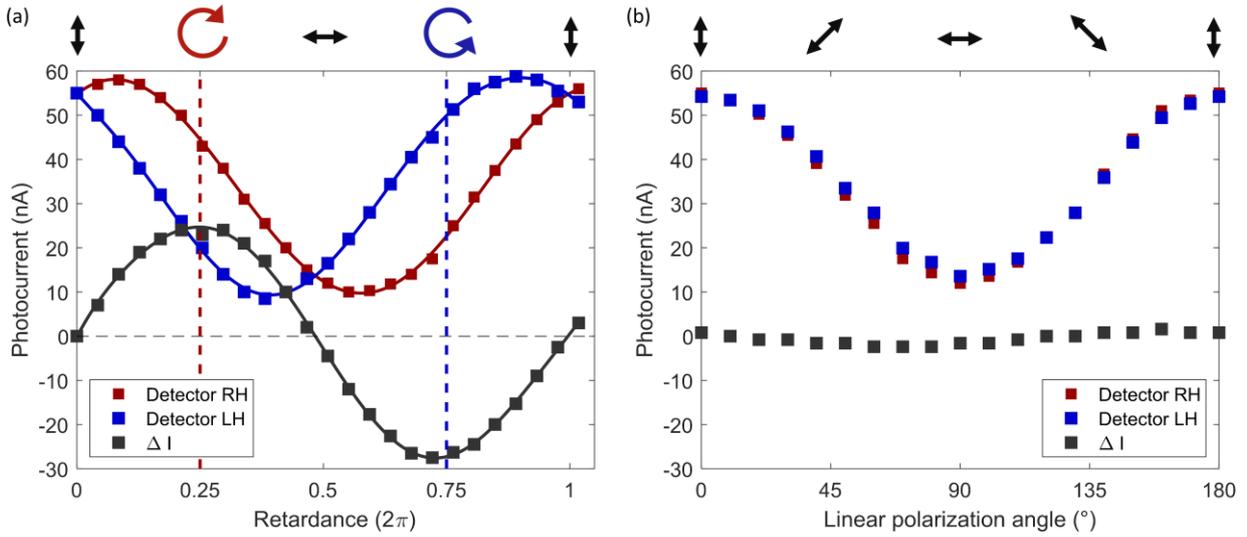

***Figure 4*** *Measured photocurrents of the individual photodetectors with the differential signal are measured for various polarizations. (a) A Soleil-Babinet variable phase retarder is used to convert linearly polarized laser radiation into left and right circularly polarized states with intermediate elliptical and an orthogonal linear polarization state. The measured photocurrents reveal that the SOI process in our device is contributing to a spin-controlled unidirectional coupling process, resulting in unbalanced photocurrents in the two detectors. The differential photocurrent signal can directly be translated to the handedness and helicity of the incident beam. (b) A half-wave plate is used to measure the photocurrents for linearly polarized light at different polarization angles. A near-zero signal in the differential photocurrent signal substantiate the fact that directionality in our device is exclusively induced by SOI.*

**TOC**

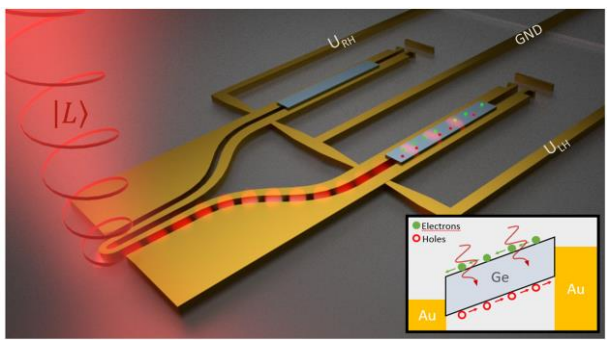